\documentclass[twocolumn,showpacs,preprintnumbers,amsmath,amssymb]{revtex4}

\usepackage{graphicx}
\usepackage{dcolumn}
\usepackage{bm}

\begin{document}

\preprint{APS/123-QED}

\title{Variation of the fundamental constants over the cosmological time: veracity of Dirac's intriguing hypothesis}
\author{Cl\'audio Nassif** and A. C. Amaro de Faria Jr*.\\
 cnassif172@gmail.com** (retired professor**), antoniocarlos@ieav.cta.br} 

 \altaffiliation{{\bf *CBPF}: Centro Brasileiro de Pesquisas F\'isicas, Rua Dr.Xavier Sigaud
 150, Urca, 22.290-180, Rio de Janeiro-RJ, Brazil.\\
 Residence address: Rua Rio de Janeiro n.1186, ap.1304, 30.160-041 (whose name of fantasy is CPFT: Centro de Pesquisas em F\'isica Te\'orica, i.e., a {\bf non-profit} name of fantasy), Belo Horizonte-MG, Brazil.\\
  {\bf *IEAv}: Instituto de Estudos Avancados, Rodovia dos Tamoios Km 099, 12220-000, S\~ao Jos\'e dos Campos-SP, Brazil.}

\date{\today}

\begin{abstract}
We investigate how the universal constants, including the fine structure constant, have varied since the early universe close to the
Planck energy scale ($E_P\sim 10^{19}$GeV) and, thus, how they have evoluted over the cosmological time related to the temperature of 
the expanding universe. According to a previous paper\cite{1}, we have shown that the speed of light was much higher close to 
the Planck scale. In the present work, we will go further, first by showing that both the Planck constant and the electron charge were 
also too large in the early universe. However, we conclude that the fine structure constant ($\alpha\cong 1/137$) has remained invariant
with the age and temperature of the universe, which is in agreement with laboratory tests and some observational data. Furthermore, we will
obtain the divergence of the electron (or proton) mass and also the gravitational constant ($G$) at the Planck scale. Thus, we will be able to verify the veracity of Dirac's belief about the existence of ``coincidences" between dimensionless ratios of sub-atomic and cosmological quantities, leading to a variation of $G$ with time, i.e., the ratio of the electrostatic to gravitational force between an electron and a proton ($\sim 10^{41}$) is roughly equal to the age of the universe divided by an elementary time constant, so that the strength of gravity, as determined by $G$, must vary inversely with time just in the approximation of lower temperature or for times very far from the early period, in order to compensate for the time-variation of the Hubble parameter ($H\sim t^{-1}$). In short, we will show the validity of Dirac's hypothesis only for times very far from the early period or $T\ll T_P$ ($\sim 10^{32}$K). 
\end{abstract}

\maketitle

\section{\label{sec:level1} Introduction}
 There are many theoretical proposals for variation of the fundamental constants of nature, including the variation of the fine structure
 constant $\alpha$\cite{2}\cite{3}\cite{4}\cite{5}\cite{6}. Furthermore, many evidences behind recent claims of spatial variation in the
 fine structure constant, due to ground-based telescopes for the observations of ion absorption lines in the light of distant
 quasars, have led to much discussion because of the controversial 
 results about how different telescopes should observe distinct spatial variations on $\alpha$\cite{7}. Variation over cosmological time
 has also been conjectured\cite{8}\cite{9}\cite{10}. In view of all this, we should be careful to investigate the veracity of such
 controversial results. For this, we will start from a significant result of the extended relativistic dynamics due to the
 presence of an isotropic background field with temperature $T$, which has been addressed in a previous Brief Report\cite{1}, where we have found 
 the dependence of the speed of light with temperature of the expanding universe. Starting from this result\cite{1}, the present work goes 
 further in order to obtain the variation of the Planck constant with temperature, the behavior of the electron charge with temperature,
 the electron (or proton) mass which has varied with the background temperature and, finally, the behavior of the gravitational constant
 at very high temperatures in the early universe and also at lower temperatures for long times, confirming Dirac's hypothesis
 ($G\sim t^{-1}$).

 Hence, we will show that the veracity of Dirac's intriguing hypothesis about the existence of ``coincidences" between dimensionless ratios 
 of sub-atomic and cosmological quantities, linking the micro and macro-universe, leads to a variation of $G$ with time, i.e., we will 
 verify that the ratio of the electrostatic force ($F_e$) to gravitational force ($F_g$) between an electron and a proton ($\sim 10^{41}$)
 is roughly equal to the age of the universe divided by an elementary time constant, so that the strength of gravity, as determined by $G$,
 must vary inversely with time just in the approximation of lower temperature or for times very far from the early period, in order to 
 compensate for the time-variation  of the Hubble parameter ($H\sim t^{-1}$). Furthermore, at the end of the last section, we will also
 verify that such a ratio of $10^{41}$, as a dimensionless number, does not vary with temperature, i.e., $F_e/F_g=F_e^{\prime}/F_g^{\prime}
 =F_e(T)/F_g(T)\sim 10^{41}$. 

 We conclude that the fine structure constant $\alpha$, as a dimensionless number as well as the ratio $F_e/F_g$, has also remained 
 invariant with the cosmic time scale (temperature). Thus, we will show that $\alpha^{\prime}=\alpha(T)=\alpha=
 q_e^2/4\pi\epsilon_0\hbar c$, where $\alpha^{-1}\approx 137.035999037(91)$\cite{11}. 
 
 A recent article on the Bayesian reanalysis of the quasar dataset\cite{12} reveals significant support for a skeptical interpretation 
in which the apparent dipole effect is driven solely by systematic errors of opposing sign inherent in measurements from the Keck and VLT 
telescopes employed to obtain the observations\cite{7}. Thus, this reanalysis leads us to question such results which show 
that the fine structure constant exhibits spatial variations. This strengthens our defense in favor of its isotropy and also its
invariance, according to Occam's razor\cite{12}. 

 It is still important to mention the observational results of J. Bahcall, W. Sargent and M. Schmidt\cite{13} who measured the fine 
structure constant in quasar $3$C$191$ and showed that its value did not vary significantly with time, giving support to our theoretical
result.  

 Although recent astrophysical data suggest that the fine structure constant $\alpha$ has increased over cosmological time, where the
combined analysis over more than 100 quasar systems has produced a value of a relative change of $\Delta\alpha/\alpha=-0.57\pm 0.10
\times 10^{-5}$, which is at the $5\sigma$ significance level\cite{14}, in contrast, we have laboratory tests that cover only a short 
time span and they have found no indications for the time-variation of $\alpha$\cite{15}. Their advantage, however, is their great
accuracy, reproducibility and unequivocal interpretation.

 \section{\label{sec:level1} Energy equation of a particle in a thermal background field}

According to the relativistic dynamics, the relativistic mass of a particle is $m=\gamma m_0$, where $\gamma=1/\sqrt{1-v^2/c^2}$ and
$m_0$ is its rest mass. On the other hand, according to Newton second law applied to its relativistic momentum, we find 
$F=dP/dt=d(\gamma m_0 v)/dt=(m_0\gamma^3)dv/dt=m_0(1-v^2/c^2)^{-3/2}dv/dt$, where $m_0\gamma^3$ represents an inertial mass 
($m_i$) that is larger than the relativistic mass $m(=\gamma m_0)$; i.e., we have $m_i>m$. 

The mysterious discrepancy between the relativistic mass $m$($m_r$) and the inertial mass $m_i$ from Newton second law is a controversial
issue\cite{16}\cite{17}\cite{18}\cite{19}\cite{20}\cite{21}\cite{22}. Actually, the Newtonian notion about inertia as the resistance to 
acceleration ($m_i$) is not compatible with the relativistic dynamics ($m_r$) in the sense that we generally cannot consider 
$\vec F= m_r\vec a$. An interesting explanation for such a discrepancy is to take into consideration the influence of an isotropic
background field\cite{1}\cite{23} that couples to the particle, by dressing its relativistic mass ($m_r$) in order to generate an effective
(dressed) mass $m^*(=m_{effective})$ working like the inertial mass $m_i(>m_r)$, in accordance with the Newtonian concept of inertia, where
we find $m^*=m_i=\gamma^2 m_r=\gamma^2 m$. In this sense, it is natural to conclude that $m^*$ has a non-local origin; i.e., it comes from
a kind of interaction with a background field connected to a universal frame\cite{23}, which is within the context of the ideas of 
Sciama\cite{24}, Schr\"{o}dinger\cite{25} and Mach\cite{26}.

If we define the new factor $\gamma^2=\Gamma$, then we write 

\begin{equation}
 m^*=\Gamma m,
\end{equation}
where $\Gamma$\cite{1} provides a non-local dynamic effect due to the influence of a universal background field on the particle moving
with speed $v$ with respect to such a universal frame\cite{25}. According to this reasoning, the particle is not completely free,
since its relativistic energy is now modified by the presence of the whole universe, namely:  

\begin{equation}
E^*= m^*c^2=\Gamma m c^2
\end{equation}

As the modified energy $E^*$ can be thought of as being the energy $E$ of the free particle plus an increment $\delta E$
of non-local origin, i.e., $E^*=\Gamma E=E+\delta E$, then let us now consider that $\delta E$ comes from a thermal background bath
due to the whole expanding universe instead of a dynamic effect ($m^*$) of a particle moving with speed $v$ in the background field,
in spite of the fact that there should be an equivalence between the dynamical and thermal effects for obtaining the modified energy.
To show this\cite{1}, we make the following assumption inside the factor $\Gamma$, namely:

\begin{equation}
\Gamma(v)=\left(1-\frac{v^2}{c^2}\right)^{-1}\equiv\Gamma(T)=\left(1-\frac{\frac{m_Pv^2}{K_B}}{\frac{m_Pc^2}{K_B}}\right)^{-1}, 
\end{equation}
from where we find $\Gamma(T)=(1-T/T_P)^{-1}$\cite{1}, $T$ being the background temperature. $T_P(=m_Pc^2/K_B\sim 10^{32}$K) is the Planck
temperature in the early universe with Planck radius $R_P\sim 10^{-35}$m. $E_P(=m_Pc^2\sim 10^{19}$GeV) is
the Planck energy and $m_P(\sim 10^{-4}$g) is the Planck mass. From the thermal approach, if $T\rightarrow T_P$, $\Gamma(T)$ diverges. 

 Now, we simply rewrite Eq.(2) as follows: 

   \begin{equation}
 E=\Gamma(T)mc^2=\frac{mc^2}{1-\frac{T}{T_P}}
   \end{equation}

It is curious to notice that the equation of Magueijo-Smolin in their doubly special relativity ($mc^2/1-E/E_P$)\cite{27} reproduces Eq.(4)
\cite{1} when we just replace $E$ by $K_BT$ and $E_P$ by $K_BT_P$ in the denominator of their equation. 

 As the factor $\Gamma(T)$ has a non-local origin and is related to the background temperature of the universe, let us admit that this 
factor acts globally on the speed of light $c$, while the well-known factor $\gamma$ acts locally on the relativistic mass of the
particle. In view of this, we should redefine Eq.(4) in the following way:

\begin{equation}
 E=[\gamma^{\prime} m_0][\Gamma(T)c^2]=\gamma^{\prime}m_0c^{\prime 2}=mc(T)^2=mc^{\prime 2},
\end{equation}
where now we have $m=\gamma^{\prime}m_0$, so that 

\begin{equation}
 \gamma^{\prime}=\frac{1}{\sqrt{1-\frac{v^2}{c^{\prime 2}}}}
\end{equation}

 And, from Eq.(5) we extract

\begin{equation}
 c^{\prime}=c(T)=\frac{c}{\sqrt{1-\frac{T}{T_P}}}, 
\end{equation}
 where $c(T)=\sqrt{\Gamma(T)}c=\gamma_{T} c$, with $\gamma_{T}=1/\sqrt{1-T/T_P}$. So the change in the speed of light is
 $\delta c=c^{\prime}-c$, i.e., $\delta c=(\gamma_{T}-1)c=(1/\sqrt{1-T/T_P}-1)c$. For $T<<T_P$, we get $\delta c\approx 0$.
  When $T=T_P$, $c^{\prime}$ has diverged. 

 From Eq.(7), we find that the speed of light was infinite in the early universe when $T=T_P$. As the universe is expanding 
and getting colder, the speed of light had been decreased to achieve $c(T)\approx c$ for $T<<T_P$. Currently we have $c(T_0)=c$,
with $T_0\approx 2.73$K. 

 In a previous work\cite{1}, basing on Eq.(7), we have shown that the speed of light in the early universe was drastically 
decreased even before the beginning of the inflationary period. Thus we were led to conclude that the usual theories of 
Varying Speed of Light (VSL) should be questioned as a possible solution for the horizon problem.  

 We should note that the variation of the speed of light with temperature of the expanding universe does not invalidate the postulate
of its invariance in the theory of special relativity, since $c^{\prime}$ given at a certain background temperature still remains 
invariant only with respect to the motion of massive particles, but not with respect to the temperature and so the cosmological time.  

\section{\label{sec:level1} Invariance of the fine structure constant with cosmological time}

From a modified relativistic dynamic of a particle moving in a space-time with an isotropic background field at a given temperature, 
it has been shown the dependence of the speed of light with temperature of the expanding universe, shown in Eq.(7). So we have found
that the energy of a particle moving under the influence of such a thermal background field is $E=\Gamma(T)mc^2=mc^{\prime2}$\cite{1}. 

Now, let us consider the energy of a photon modified by the presence of a given background temperature. So we write

\begin{equation}
E=mc^{\prime 2}=p^{\prime}c^{\prime},
\end{equation}
where $p^{\prime}=mc^{\prime}=mc(T)=m\gamma_Tc$, which represents the modified momentum of the photon. 

On the other hand, we already know that the energy of a photon is $E=h\nu=\hbar w$, where $\hbar=h/2\pi$ and $w=2\pi\nu=2\pi c/\lambda$,
$\lambda$ being the wavelength of the photon and $\nu(=c/\lambda)$ being its frequency. Now, if we consider this energy modified 
by the background temperature, we find 

\begin{equation}
E=\hbar^{\prime}w^{\prime}=h^{\prime}\nu^{\prime}=h^{\prime}\frac{c^{\prime}}{\lambda},
\end{equation}
where $\nu^{\prime}=c^{\prime}/\lambda$. 

 Comparing Eq.(8) to Eq.(9), we write
\begin{equation}
E=mc^{\prime 2}=\lambda^{-1}h^{\prime}c^{\prime},  
\end{equation}

or simply 

\begin{equation}
p^{\prime}=mc^{\prime}=\lambda^{-1}h^{\prime}=\lambda^{-1}h(T),  
\end{equation}
which represents the de-Broglie equation for the photon in the presence of the background temperature. 

According to Eq.(7), Eq.(11) is written as $m(\gamma_Tc)=\lambda^{-1}h^{\prime}$. Since both the wavelength $\lambda$ and the 
relativistic mass $m$ of the photon are not corrected with temperature, only the universal constants $c$ and $h$, as global quantities, 
are influenced by the background temperature, so that we have $c^{\prime}(=\gamma_Tc)$ and $h^{\prime}[=h(T)]$ in Eq.(11). Thus,
from Eq.(11), we conclude that the Planck constant should be corrected in the same way of the speed of light in Eq.(7). So we find

\begin{equation}
 h^{\prime}=h(T)=\gamma_Th=\frac{h}{\sqrt{1-\frac{T}{T_P}}}, 
\end{equation}
or else $\hbar^{\prime}=\hbar(T)=\gamma_T\hbar$, with $\hbar=h/2\pi$. So, from Eq.(11), we find $p^{\prime}=m(\gamma_Tc)=
\lambda^{-1}(\gamma_Th)\Rightarrow\lambda=h/mc$, in such a way to preserve the de-Broglie equation and, thus, the wavelength
 (frequency) of the photon. In the early universe, when $T=T_P$, $h^{\prime}$ has diverged. 

It is known that $c^2=1/\mu_0\epsilon_0$, where $\mu_0$ is the magnetic permeability of vacuum and $\epsilon_0$ is the electric
permittivity of vacuum. Thus, based on Eq.(7), by correcting this Maxwell relation with temperature of the universe, we write 

\begin{equation}
c^{\prime 2}=\frac{1}{\mu_0^{\prime}\epsilon_0^{\prime}}=\frac{c^2}{1-\frac{T}{T_P}}=\frac{1}{\mu_0\epsilon_0\left(1-\frac{T}{T_P}\right)},
\end{equation}
from where we extract 

\begin{equation}
\mu_0^{\prime}=\mu_0(T)=\mu_0\sqrt{1-\frac{T}{T_P}}, 
\end{equation}

and 

\begin{equation}
\epsilon_0^{\prime}=\epsilon_0(T)=\epsilon_0\sqrt{1-\frac{T}{T_P}}, 
\end{equation}
since the electric ($\epsilon$) and magnetic ($\mu$) aspects of radiation are in equal-footing.

Based on the electric interaction energy $\Delta E_e$ between two point-like electrons separated by a certain distance $r$, we write

\begin{equation}
\Delta E_e=\Delta m_ec^2=\frac{e^2}{r}=\frac{q_e^2}{4\pi\epsilon_0r},
\end{equation}
where $e^2=q_e^2/4\pi\epsilon_0$. $\Delta E_e$($=\Delta m_ec^2$) is the relativistic representation for such an interaction energy of
electric origin, which decreases to zero when $r\rightarrow\infty$, i.e., we have $\Delta m_e\rightarrow 0$ for $r\rightarrow\infty$. 

Now, by correcting Eq.(16) due to the presence of a thermal background field according to Eq.(7), we write

\begin{equation}
\Delta m_ec^{\prime 2}=\Delta m_ec(T)^2=\frac{\Delta m_ec^2}{1-\frac{T}{T_P}}=\frac{e^{\prime 2}}{r}=
\frac{q_e^{\prime 2}}{4\pi\epsilon_0^{\prime}r},
\end{equation}
from where we get

\begin{equation}
 e^{\prime 2}=e^2(T)=\frac{e^2}{1-\frac{T}{T_P}}, 
\end{equation}

or else

\begin{equation}
 \frac{q_e^{\prime 2}}{4\pi\epsilon_0^{\prime}}=\frac{q_e^2}{4\pi\epsilon_0\left(1-\frac{T}{T_P}\right)}
\end{equation}

Inserting Eq.(15) into Eq.(19), we find

\begin{equation}
 \frac{q_e^{\prime 2}}{4\pi\epsilon_0\sqrt{1-\frac{T}{T_P}}}=\frac{q_e^2}{4\pi\epsilon_0\left(1-\frac{T}{T_P}\right)}, 
\end{equation}
which implies

\begin{equation}
 q_e^{\prime 2}=\gamma_Tq_e^2=\frac{q_e^2}{\sqrt{1-\frac{T}{T_P}}},
\end{equation}

or

\begin{equation}
 q_e^{\prime}=q_e(T)=\frac{q_e}{\sqrt[4]{1-\frac{T}{T_P}}}
\end{equation}

\begin{figure}
\includegraphics[scale=0.22]{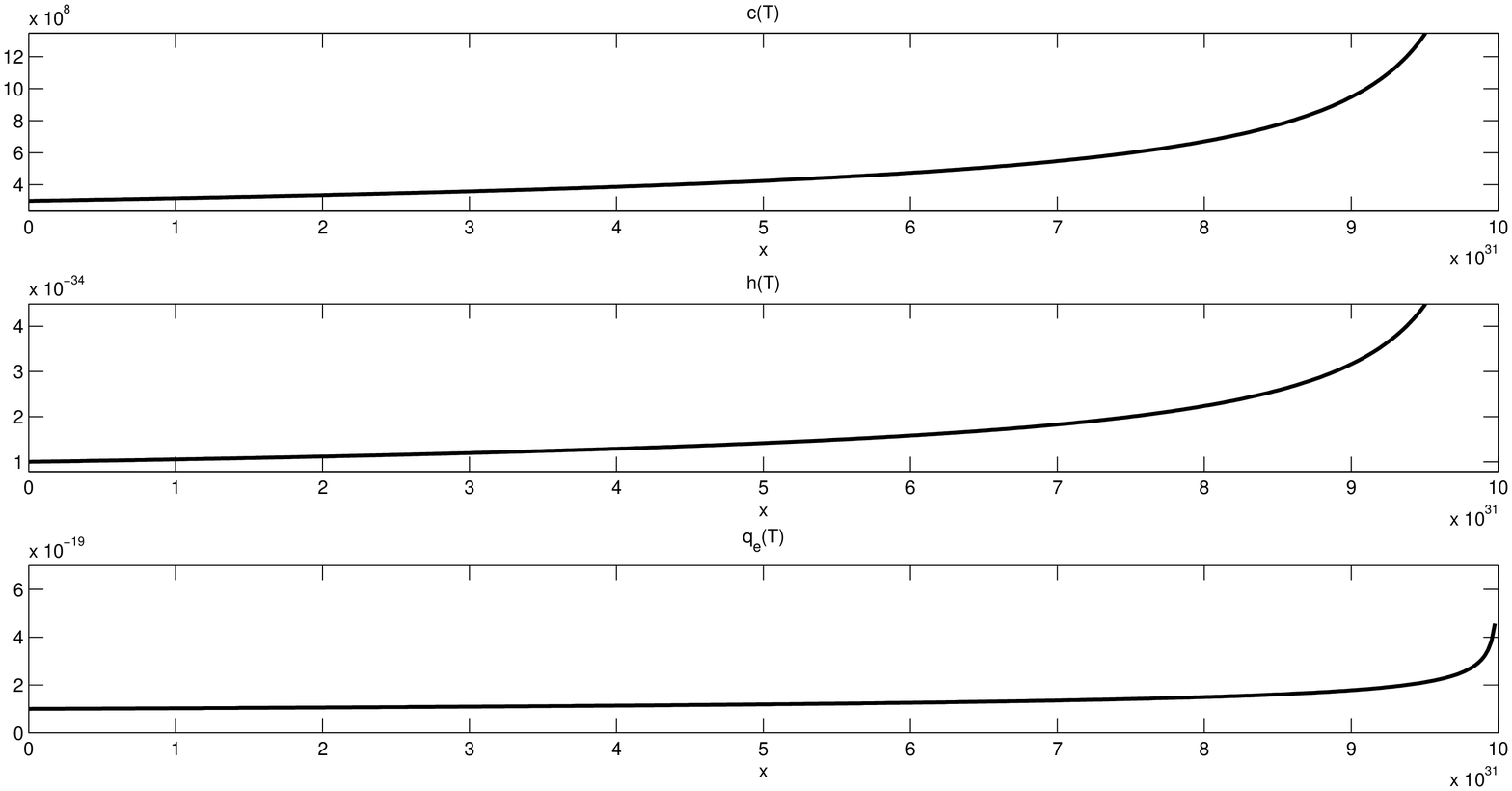}
\caption{{\it The three graphs above provide variations in $c$, $h$ and $q_e$ from top to bottom respectively, where there are divergences at Planck scale ($\sim 10^{32}$K).}}
\end{figure}

The fine structure constant without temperature is $\alpha=e^2/\hbar c=q_e^2/4\pi\epsilon_0\hbar c=q_e^2\mu_0 c/2h$. Now, by taking into account a given temperature of the expanding universe, we have

\begin{equation}
\alpha(T)=\alpha^{\prime}=\frac{e^{\prime 2}}{\hbar^{\prime}c^{\prime}}=\frac{q_e^{\prime 2}}{4\pi\epsilon_0^{\prime}
\hbar^{\prime}c^{\prime}}=\frac{q_e^{\prime 2}\mu_0^{\prime}c^{\prime}}{2h^{\prime}}
\end{equation}

Finally, by inserting Eq.(7)($c^{\prime}$), Eq.(12)($\hbar^{\prime}$) and Eq.(18)($e^{\prime 2}$) into Eq.(23), or by inserting
$c^{\prime}$, $\hbar^{\prime}$, $\epsilon_0^{\prime}$[Eq.(15)] and $q_e^{\prime 2}$[Eq.(21)] into Eq.(23), or even by inserting
$c^{\prime}$, $h^{\prime}$, $\mu_0^{\prime}$[Eq.(14)] and $q_e^{\prime 2}$[Eq.(21)] into Eq.(23), we find

\begin{equation}
\frac{e^{\prime 2}}{\hbar^{\prime}c^{\prime}}=\frac{e^2}{\hbar c}, 
\end{equation}

or

\begin{equation}
\frac{q_e^{\prime 2}}{4\pi\epsilon_0^{\prime}\hbar^{\prime}c^{\prime}}=\frac{q_e^2}{4\pi\epsilon_0\hbar c} 
\end{equation}

or

\begin{equation}
\frac{q_e^{\prime 2}\mu_0^{\prime}c^{\prime}}{2h^{\prime}}=\frac{q_e^2\mu_0 c}{2h}, 
\end{equation}

that is, 

\begin{equation}
\alpha^{\prime}=\alpha\approx\frac{1}{137.035999037(91)},
\end{equation}
which reveals to us the invariance of the fine structure constant with temperature of the expanding universe and thus its invariance over cosmological time.

\begin{figure}
\includegraphics[scale=0.22]{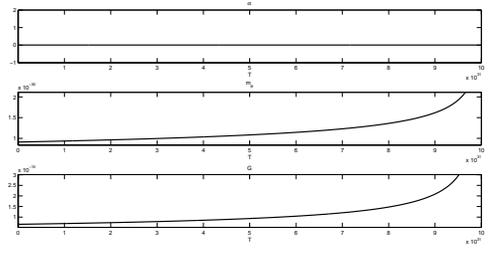}
\caption{{\it The first graph above shows the invariance of the fine structure constant with temperature of the universe. The other two graphs below show the divergences of $m_e$ and $G$ close to the Planck scale ($\sim 10^{32}$K).}}
\end{figure}

\section{\label{sec:level1} Variation of the gravitational constant in the early universe}

Now, by considering the very low energy of gravitational interaction $\Delta E_g$ between two point-like electrons separated by a
certain distance $r$, we write

\begin{equation}
\Delta E_g=\Delta m_gc^2=\frac{Gm_e^2}{r},
\end{equation}
where $\Delta E_g$($=\Delta m_gc^2$) is the relativistic representation for such an interaction energy of gravitational origin, which is
much lower than the interaction energy of electric origin, i.e., we have $\Delta m_g\ll\Delta m_e$ ($\Delta E_g\ll\Delta E_e$), where
$\Delta m_g$ also decreases to zero when $r\rightarrow\infty$, i.e., we have $\Delta m_g\rightarrow 0$ for $r\rightarrow\infty$. 

Now, by correcting Eq.(28) due to the presence of the thermal background field according to Eq.(7), we write

\begin{equation}
\Delta m_gc^{\prime 2}=\Delta m_gc(T)^2=\frac{\Delta m_gc^2}{1-\frac{T}{T_P}}=\frac{G^{\prime}m_e^{\prime 2}}{r},
\end{equation}
from where we get

\begin{equation}
 G^{\prime}m_e^{\prime 2}=G(T)m_e^2(T)=\frac{Gm_e^2}{1-\frac{T}{T_P}}, 
\end{equation}
and from where we extract separately: 

\begin{equation}
 m_e^{\prime}=m_e(T)=\frac{m_e}{\sqrt[4]{1-\frac{T}{T_P}}}
\end{equation}

and 

\begin{equation}
 G^{\prime}=G(T)=\frac{G}{\sqrt{1-\frac{T}{T_P}}}, 
\end{equation}
where we have admitted that $m_e$ (Eq.31) and $q_e$ (Eq.22) are on equal footing with respect to their variation with temperature, i.e.,
the ratio $m_e/q_e$ has been remained invariant with the cosmological time, since the dimensionless number $\alpha_g=Gm_e^2/\hbar c$
should be also invariant like $\alpha=e^2/\hbar c$. Thus we can realize that $m_e$ and $q_e$ diverged likewise close to the Planck 
temperature. 

According to Eq.(32), $G$ diverges for $T=T_P$. 

\subsection{\label{sec:level1} Veracity of Dirac's intriguing hypothesis}

In analogous way to the reasoning above for the gravitational interaction between two electrons, now if we consider the gravitational 
 interaction between two protons in the presence of such a thermal background field, we find: 

\begin{equation}
 m_p^{\prime}=m_p(T)=\frac{m_p}{\sqrt[4]{1-\frac{T}{T_P}}},
\end{equation}
where $m_p$ is the proton mass. 

In his intriguing hypothesis, Dirac\cite{28} noted that, for some unexplained reason, the ratio of the electrostatic to gravitational force 
between an electron and a proton is roughly equal to the age of the universe divided by an elementary time constant, implying that 

\begin{equation}
 \frac{F_e}{F_g}=\frac{\alpha hc}{Gm_pm_e}\sim\frac{m_pc^2}{Hh}\sim 10^{41}, 
\end{equation}
where $H$ is the Hubble parameter. And, it is easy to verify that the ratio $F_e/F_g$ remains invariant with temperature, i.e., 
$F_e/F_g=F_e^{\prime}/F_g^{\prime}=\alpha^{\prime}h^{\prime}c^{\prime}/G^{\prime}m_p^{\prime}m_e^{\prime}\sim 10^{41}$. 

 Eq.(34) led Dirac to argue that the strength of gravity, as determined by $G$, must vary inversely with time, to compensate 
for the time-variation of the Hubble parameter $H$. Indeed, we will show that this Dirac's hypothesis can be confirmed within a certain 
approximation only for $T\ll T_P$, when we obtain $H^{\prime}[=H(T)]$ in such a way as to preserve the coincidence and invariance of the
ratio of $10^{41}$ for any temperature in Eq.(34)\cite{29}\cite{30}. To do that, we simply write 

\begin{equation}
\frac{m_p^{\prime}c^{\prime2}}{H^{\prime}h^{\prime}}\sim 10^{41} 
\end{equation}

In order to preserve the invariance of the ratio of $10^{41}$ with temperature in Eq.(35), we first insert Eq.(7) ($c^{\prime}$), 
Eq.(12) ($h^{\prime}$) and Eq.(33) ($m_p^{\prime}$) into Eq.(35) and, after, we should perform the calculations, such that we must find 

\begin{equation}
H^{\prime}=H(T)=H\left(1-\frac{T}{T_P}\right)^{-\frac{3}{4}}\approx H\left(1+\frac{3}{4}\frac{T}{T_P}\right), 
\end{equation}
which is only valid when $T\ll T_P$, becoming a linear function of the form $H^{\prime}= H + aT$, $a$ being a constant. 

In order to make the approximation given in Eq.(36) to be consistent with the time-variation of the Hubble parameter, i.e., 
$H\sim t^{-1}$, which is also an approximation valid just for long times, thus, we must extract the following approximation,
namely $T\sim t^{-1}$ for $T\ll T_P$. So, finally, by doing this, we can write Eq.(32) as follows: 

\begin{equation}
G^{\prime}=G(T)=G\left(1-\frac{T}{T_P}\right)^{-\frac{1}{2}}\approx G\left(1+\frac{1}{2}\frac{T}{T_P}\right), 
\end{equation}
when $T\ll T_P$, such that we get a linear function $G^{\prime}= G + bT$, $b$ being a constant. Hence, since $T\sim t^{-1}$, we 
can simply write $G\sim t^{-1}$, confirming Dirac's hypothesis.

\end{document}